# Physics-guided inverse design of Co-based superalloys using machine learning and multi-objective optimization for enhanced γ′ solvus temperature

Prashil S. Joshi[1*]

(1) Prashil S. Joshi[1] (***Corresponding author**)

E-mail: prashiljoshi@iisc.ac.in

Affiliation(s): [1]Department of Materials Engineering, Indian Institute of Science, Bengaluru,

Karnataka - 560012, India.

ORCID: orcid.org/0009-0009-6727-5483

**Abstract:**

The discovery of next-generation Co-based superalloys with improved high-temperature stability is hindered by the vast compositional design space and the complex interactions among alloying elements that govern γ′ phase stability. This study presents a physics-informed machine learning framework for the inverse design of Co-based superalloys that achieve a higher γ′ solvus temperature while accounting for alloy density. Four significant descriptors reflecting atomic size mismatch (δMV), mixing enthalpy ($\Delta H_m$), electronegativity mismatch (δEN), and valence electron concentration mismatch (δVEC) were used to understand the chemistry that influences phase stability. Various regression algorithms were assessed through leave-one-out cross-validation, revealing that Gaussian Process Regression (GPR) achieved the best predictive performance ($R^2$ = 0.932, RMSE = 34.7 °C and MAE = 25.9 °C). The probabilistic nature of GPR enabled uncertainty-aware Bayesian Optimization for effective exploration of the uncharted composition space. Concurrently, Non-dominated Sorting Genetic Algorithm II (NSGA-II) and Genetic Algorithms were combined to enhance the γ′ solvus temperature while simultaneously reducing alloy density. The optimization framework produced numerous Co-based alloy compositions not previously studied, with predicted γ′ solvus temperatures ranging from 1248 to 1353 °C, and several candidates nearing or surpassing the maximum value observed in the experimental dataset. Descriptor importance analysis revealed that atomic size mismatch and mixing enthalpy are the key factors influencing γ′ stability, and principal component analysis confirmed that the proposed alloys lie within chemically relevant regions of descriptor space. The optimized formulations consistently maintained elevated levels of Co and W while permitting adjustments in secondary alloying additions to achieve a balance between thermal stability and density. The suggested physics-informed inverse design approach offers an effective method for accelerating the identification of high-performance Co-based superalloys and can be readily adapted for the design of various advanced structural alloys.

## 1. Introduction

Nickel (Ni)-based superalloys, reinforced by a significant volume fraction of the ordered, $L1_2$-structured γ′-$Ni_3Al$ precipitate coherently integrated within a disordered FCC γ matrix, have been the preferred material for more than fifty years for the hottest structural elements of aircraft and land-based gas-turbine engines, such as turbine blades, vanes, and discs [1–3]. Their achievements arise from a unique blend of elevated-temperature strength, creep and fatigue resistance, and microstructural stability, all of which originate from the coherent γ/γ′ two-phase microstructure and can be further customized by incorporating refractory elements such as W, Ta, Mo, and Re [1,3]. Nevertheless, the ongoing drive for increased turbine entry temperatures, driven by the related improvements in thermal efficiency and decreases in specific fuel consumption and emissions, has pushed Ni-based superalloys near their fundamental temperature limits, as their highest service temperature is fundamentally constrained by the melting point of Ni and by γ′ solvus temperatures that seldom go beyond approximately 1150–1200 °C in even the most advanced commercial alloys [1,2]. This has driven an ongoing quest for alternative or complementary high-temperature structural materials that can broaden the operational range of gas-turbine hot sections beyond the capabilities of Ni-based superalloys. Cobalt (Co), located adjacent to Ni in the periodic table, has been recognized for years as a favorable base element for high-temperature alloys due to its elevated melting point (1495 °C, compared to 1455 °C for Ni) and its generally enhanced resistance to hot corrosion [4]. Historically, traditional Co-based superalloys depended on solid-solution strengthening of the FCC γ matrix and the formation of $M_{23}C_6$- and MC-type carbides at grain boundaries, as there were no known beneficial ordered intermetallic strengthening phases similar to γ′-$Ni_3Al$ in Co-rich systems [4,5]. As a consequence, despite their superior melting point and corrosion resistance, traditional Co-based alloys exhibited markedly lower high-temperature strength and creep resistance than γ/γ′ Ni-based superalloys and were consequently relegated to lower-stress, corrosion-dominated applications such as combustor liners and vanes rather than the highest-stress rotating components [4,5].

The landscape shifted significantly in 2006, when [4] found that a ternary $L1_2$-ordered γ′-$Co_3$(Al, W) phase could be stabilised within the Co-Al-W system, resulting in a γ/γ′ two-phase microstructure remarkably similar to that of Ni-based superalloys, yet relying on a Co-rich matrix. This finding instantly opened up the possibility of a completely new category of γ′-strengthened

Co-based superalloys that could merge the elevated melting point and exceptional hot-corrosion resistance of Co with the coherent-precipitate strengthening approach that was previously exclusive to Ni-based alloys [4,6]. In the subsequent years, a sizable collection of experimental and computational studies has aimed to elucidate the thermodynamics, phase stability, and mechanical properties of $\gamma'$-$Co_3(Al,W)$-based alloys, as well as to broaden the initial ternary system to include higher-order, industrially significant compositions featuring Ni, Cr, Ta, Ti, Mo, and Nb, among various other elements [5–11]. This body of research has highlighted both the potential and difficulties of the new alloy category: several Co-based $\gamma'$ alloys demonstrate $\gamma'$ solvus temperatures surpassing 1200 °C while maintaining densities comparable to or lower than second-generation Ni-based single-crystal superalloys, making them viable options for weight-sensitive, high-temperature aerospace uses [6,11]; concurrently, the initial binary $\gamma'$-$Co_3(Al,W)$ phase is thermodynamically metastable compared to $Co_3W$ and other competing phases at standard operating temperatures, and obtaining a sufficiently broad and stable $\gamma/\gamma'$ two-phase region, along with satisfactory oxidation resistance, has typically necessitated careful multi-element alloying approaches, with the best combinations remaining unclear beforehand [5,7–10].

A key characteristic in Co-based superalloys is the $\gamma'$ solvus temperature ($T_{\gamma'}$), which indicates the temperature at which the ordered $\gamma'$-$Co_3(Al,W)$ precipitates dissolve into the disordered $\gamma$ matrix. As the strengthening effect of the $\gamma'$ phase, due to coherency strengthening, order strengthening, and dislocation motion resistance, diminishes above this temperature, $T_{\gamma'}$ essentially sets the maximum operating temperature for these alloys, similar to its function in Ni-based superalloys [1,4,5]. Optimizing $T_{\gamma'}$ is quite difficult due to the complex, strongly nonlinear interactions among various alloying elements. Refractory elements like W and Ta significantly enhance $\gamma'$ stability and raise $T_{\gamma'}$, but also increase alloy density, whereas Cr primarily improves oxidation resistance with a milder effect on phase stability. Various alloying additions, such as Ni, Ti, Mo, and Nb, alter the $\gamma/\gamma'$ partitioning behavior and precipitation traits in ways that are challenging to predict solely from first-principles calculations. Thus, determining alloy compositions that concurrently enhance $T_{\gamma'}$, reduce density, and inhibit the formation of unwanted secondary phases represents a complex, multi-objective optimization challenge [5–11]. Historically, investigating this design space has depended on iterative experimental trial-and-error backed by CALPHAD (CALculation of PHAse Diagrams) thermodynamic modeling, where potential compositions are suggested based on physical intuition and previous understanding, synthesized, characterized, and later improved.

Even though CALPHAD-guided alloy design has facilitated the creation of numerous high-performance Co-based superalloys, its predictive ability is still constrained by the presence of dependable thermodynamic databases for intricate multicomponent systems, the computational expense of navigating a high-dimensional compositional space, and the reliance on expert opinion during candidate selection [8–11]. These constraints have sparked increased fascination with materials design driven by machine learning (ML), in which predictive models learn composition-property relationships directly from experimental or computational datasets [12,13]. Specifically, probabilistic surrogate models like Gaussian Process Regression (GPR), combined with Bayesian optimization, have become effective tools for uncertainty-informed inverse design by concurrently leveraging areas anticipated to exhibit enhanced properties while probing uncertain regions of the compositional space via acquisition functions such as Expected Improvement [14–17]. These methods have proven effective for the rapid discovery of NiTi shape-memory alloys, high-entropy alloys, ferroelectric perovskites, precipitation-strengthened alloys, and various other advanced material systems, demonstrating their ability to significantly reduce experimental work while effectively exploring extensive compositional design spaces[10,18–23].

In the realm of $\gamma'$-strengthened Co-based superalloys, numerous recent investigations have showcased the promise of machine learning (ML) for expedited alloy development. [12] utilized machine learning alongside multi-property optimization to discover Co-based superalloys exhibiting enhanced $\gamma'$ solvus temperature, density, oxidation resistance, and processing traits from an extensive compositional range, which was subsequently verified through experimental testing of the predicted alloys. [9] combined machine learning with CALPHAD calculations to create tungsten-free Co-based superalloys with an enlarged $\gamma/\gamma'$ two-phase region, whereas [13] established a specific machine learning model to predict the $\gamma'$ solvus temperature of $L1_2$-strengthened Co-based superalloys based solely on alloy composition. Overall, these studies reveal that ML-supported methods can greatly accelerate alloy discovery compared to traditional CALPHAD-based or intuition-led techniques. Nonetheless, numerous significant obstacles persist. Numerous current predictive models rely on empirical or learned features, offering limited physical interpretability even though they achieve high predictive accuracy. Conversely, descriptors based on physics that were initially developed for high-entropy alloys, such as atomic size mismatch, mixing enthalpy, electronegativity mismatch, and valence electron concentration mismatch, offer analytically calculable values with clear physical meaning [24–31]. Moreover, since both the solvus temperature of γ′ and density are essential design goals for aerospace uses, optimization should preferably be set up as a true multi-objective issue instead of merging various properties into a randomly weighted objective function [6,11,32]. Moreover, when GPR is used for Bayesian optimization, accurate uncertainty estimation is crucial since the Expected Improvement acquisition function relies directly on properly calibrated predictive uncertainties [15–17].

In this study, a machine learning framework that incorporates physics and accounts for uncertainty is created for the inverse design of γ′-reinforced Co-based superalloys to overcome these challenges. Four significant physical descriptors—molar volume mismatch (δMV), mixing enthalpy (ΔHm), electronegativity mismatch (δEN), and valence electron concentration mismatch (δVEC), are utilized as the exclusive input features for predicting the γ′ solvus [27–31,33]. Five regression algorithms are systematically evaluated through leave-one-out cross-validation, and the top-performing GPR model undergoes uncertainty calibration validation before its integration with Bayesian Optimization, Non-dominated Sorting Genetic Algorithm II (NSGA-II) multi-objective optimization, and a Genetic Algorithm for generating candidates [15–17,32,34]. Candidate compositions are additionally restricted by a physics-informed descriptor filter, while Principal Component Analysis and k-means clustering are used to confirm that the proposed alloys remain within physically relevant regions of descriptor space and yield compositionally varied candidates [35–37]. The suggested framework identifies 10 Co-based superalloys that have not been studied before, with estimated γ′ solvus temperatures ranging from 1248 to 1353 °C, many of which exceed the highest values in the experimental dataset and span a range of alloy densities suitable for various engineering applications.

## 2. Methodology

### 2.1 Experimental dataset

An experimental dataset comprising γ′-strengthened Co-based superalloys was compiled from published literature including the compositions and γ′ solvus temperatures reported by [38]. Alloy compositions span the Co–Ni–Al–Cr–W–Ta–Ti–Mo–Nb system and were represented using elemental atomic percentages. Alloy density was calculated using the rule of mixtures and subsequently employed as a secondary optimization objective.

### 2.2 Physics-guided descriptor calculation

Four physically meaningful descriptors were calculated from the alloy compositions: molar volume mismatch (δMV), mixing enthalpy (ΔHm), electronegativity mismatch (δEN), and valence electron concentration mismatch (δVEC). These descriptors capture lattice distortion, chemical interactions, electronic structure, and compositional complexity governing $\gamma'$ phase stability, while providing an interpretable representation of alloy chemistry.

### 2.3 Machine learning models

Five regression algorithms, namely Linear Regression, Ridge Regression, Random Forest, XGBoost, and Gaussian Process Regression (GPR), were evaluated to predict the $\gamma'$ solvus temperature. Model performance was assessed using leave-one-out cross-validation (LOOCV). The coefficient of determination ($R^2$), root mean square error (RMSE) and mean absolute error (MAE) were used to compare predictive performance.

$$R^2 = 1 - \frac{\Sigma(yi - \hat{y}i)^2}{\Sigma(yi - \bar{y})^2} \qquad (1.1)$$

$$RMSE = \sqrt{\left[\left(\frac{1}{n}\right)\Sigma(yi - \hat{y}i)^2\right]} \qquad (1.2)$$

$$MAE = \left(\frac{1}{n}\right)\Sigma|yi - \hat{y}i| \qquad (1.3)$$

### 2.4 Gaussian Process Regression

Gaussian Process Regression was adopted as the primary surrogate model owing to its ability to provide both accurate predictions and predictive uncertainty. The probabilistic nature of GPR enables uncertainty-aware optimization and supports Bayesian optimization through the Expected Improvement acquisition function.

$$f(x) \sim GP\big(m(x), k(x, x')\big) \qquad (1.4)$$

where m(x) is the mean function and k is the covariance kernel. A radial basis function kernel was employed:

$$k(x,x') = \sigma f^2 exp\left(-\frac{\left||x-x'|\right|^2}{2l^2}\right) \quad (1.5)$$

The predictive mean and variance are

$$\mu^* = K^{*^{\mathrm{T}}}(K+\sigma n^2 I)^{-1} y \quad (1.6)$$

$$\sigma^{*^2} = K^{**} - K^{*^{\mathrm{T}}}(K+\sigma n^2 I)^{-1} K^* \quad (1.7)$$

### 2.5 Bayesian Optimization

Bayesian Optimization was performed using the calibrated GPR surrogate model. The Expected Improvement acquisition function was employed to balance exploration of uncertain regions and exploitation of promising compositions, thereby efficiently identifying candidate alloys with enhanced predicted γ′ solvus temperatures.

The Expected Improvement (EI) acquisition function,

$$EI(x) = (\mu - f_{best} - \xi)\Phi(Z) + \sigma\varphi(Z) \quad (1.8)$$

where

$$Z = \frac{\mu - f_{best} - \xi}{\sigma} \quad (1.9)$$

Here μ and σ are the GPR predictive mean and standard deviation, Φ and φ denote the normal cumulative and probability density functions, respectively, and $f_{best}$ is the best observed γ′ solvus temperature.

### 2.6 Multi-objective optimization using NSGA-II

NSGA-II was employed to maximize γ′ solvus temperature and minimize alloy density simultaneously. Candidate solutions were ranked using Pareto dominance, while crowding-distance selection preserved diversity along the Pareto front, yielding a family of optimal trade-off compositions.

### 2.7 Genetic Algorithm

A Genetic Algorithm was implemented as an additional global optimization strategy using the GPR surrogate model as the objective evaluator.

Population evolution through selection, crossover, and mutation enabled efficient exploration of the compositional design space.

### 2.8 Physics-guided candidate filtering

Candidate alloys generated by the optimization algorithms were screened using descriptor bounds derived from experimentally validated high-performing alloys. For each descriptor d ∈ {δMV, ΔHm, δEN, δVEC}, a candidate x was retained only if,

$$\mu_{d,top} - k \cdot \sigma_{d,top} \leq d(x) \leq \mu_{d,top} + k \cdot \sigma_{d,top} \quad (1.10)$$

where $\mu_{d,top}$ and $\sigma_{d,top}$ are the mean and standard deviation of the descriptor $d$ among the top-quartile training alloys, and $k$ is a tolerance factor.

This physics-guided filtering eliminated chemically unrealistic compositions while retaining candidates within physically meaningful regions of descriptor space.

### 2.9 Principal Component Analysis

Principal Component Analysis (PCA) was employed to visualize the descriptor space and verify that the optimized alloys remained within the distribution of experimentally reported compositions.

For the standardised, mean-centred n × 4 descriptor matrix X, PCA computes the eigen-decomposition of the descriptor covariance matrix:

$$C = \left(\frac{1}{n-1}\right) X^{T}X = V\Lambda V^{T} \quad (1.11)$$

where V is the matrix of eigenvectors (loadings), and Λ is the diagonal matrix of eigenvalues, ordered from largest to smallest. Projected scores are obtained as T = XV, with the first two columns of T giving the PC1 and PC2 coordinates plotted for each alloy. The fraction of total descriptor variance explained by the k-th principal component is

$$Explained\ variance\ ratio = \lambda_k \ / \ \Sigma_j \lambda_j \quad (1.12)$$

Alloys (training or proposed) whose PC1–PC2 coordinates fall within the envelope of the training-data distribution are interpreted as physically grounded, controlled extrapolations rather than compositions lying outside chemically meaningful regions of descriptor space.

### 2.10 K-means clustering

K-means clustering was applied to the optimized candidate pool to identify compositionally diverse solutions.

For a chosen number of clusters k = 10, the algorithm partitions the pooled candidates into disjoint sets $S = \{S_1,\ldots,S_k\}$ to minimise the within-cluster sum of squared distances to each cluster's centroid:

$$arg\ min_S \ \sum_{k}^{j=1} \sum_{x \in S_j} \left\| x - \mu_j \right\|^2 \quad (2.18)$$

where $\mu_j$ is the centroid (mean descriptor vector) of cluster $S_j$. The algorithm proceeds iteratively: (i) assign each candidate to the nearest centroid, (ii) recompute each centroid as the mean descriptor vector of the candidates currently assigned to it and repeat until the assignments no longer change (convergence). Representative alloys were selected from each cluster, avoiding redundant compositions while maximizing chemical diversity.

**2.11 Overall workflow**

The overall workflow comprised experimental dataset compilation, descriptor calculation, machine learning model development, uncertainty calibration, Bayesian optimization, NSGA-II optimization, Genetic Algorithm search, physics-guided filtering, PCA visualization, and k-means clustering to obtain ten optimized Co-based superalloy compositions for future experimental validation.

## 3. Results & Discussions

### 3.1 Descriptor–property relationships

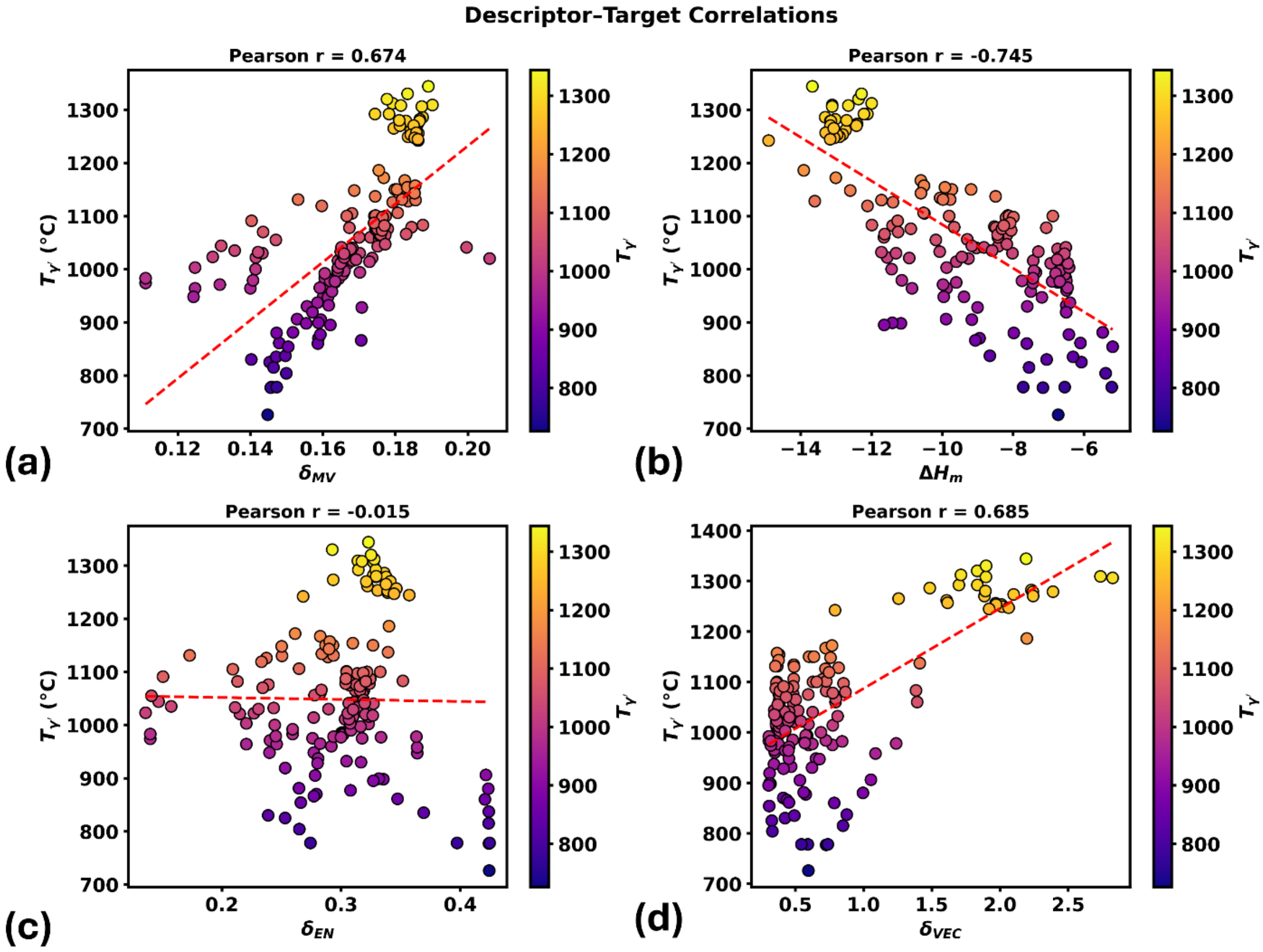

***Figure 1.*** *Correlation between the selected physics-informed descriptors and γ' solvus temperature for the experimental Co-based superalloy dataset. Scatter plots illustrate the relationships between (a) atomic size mismatch (δMV), (b) mixing enthalpy (ΔHm), (c) electronegativity mismatch (δEN), and (d) valence electron concentration mismatch (δVEC) with the experimentally measured γ' solvus temperature. Pearson correlation coefficients (r) are indicated for each descriptor.*

Figure 1 illustrates the connections between the chosen physicochemical descriptors and the γ′ solvus temperature. The four descriptors were chosen to represent different facets of alloy chemistry, encompassing lattice distortion (δMV), chemical bonding ($\Delta H_m$), electronegativity mismatch (δEN), and electronic structure (δVEC). Of the descriptors studied, the molar volume mismatch (δMV) showed a significant positive correlation with the γ′ solvus temperature (Pearson's r = 0.674), suggesting that alloys with larger atomic size mismatches typically demonstrate higher γ′ solvus temperatures. Increased lattice distortion enhances solid-solution strengthening and alters the partitioning behavior of refractory elements, thereby stabilizing the ordered γ′ precipitate. In contrast, the mixing enthalpy ($\Delta H_m$) showed the most significant negative correlation with the target property (r = –0.745). Stronger chemical interactions between constituent elements correspond to more negative values of $\Delta H_m$, promoting the formation of ordered phases and enhancing γ′ stability. This observation aligns with the recognized role of chemical bonding in influencing precipitate stability in Co-based superalloys. The mismatch in valence-electron concentration (δVEC) showed a strong positive correlation (r = 0.685), indicating that the electronic structure significantly influences γ′ stability. Alterations in the local electronic environment impact phase stability, ordering preference, and elemental distribution, consequently influencing the γ′ solvus temperature. Conversely, electronegativity mismatch (δEN) demonstrated minimal linear correlation with the desired property (r = –0.015). While δEN by itself does not seem to control γ′ solvus behavior, it could still play a role through nonlinear interactions with other descriptors, which can be effectively modeled using nonlinear machine learning techniques. In summary, the analysis of descriptors indicates that the γ′ solvus temperature is determined by the combined effects of atomic size mismatch, chemical interactions, and electronic structure, rather than by any individual descriptor alone.

### 3.2 Comparison of machine learning models

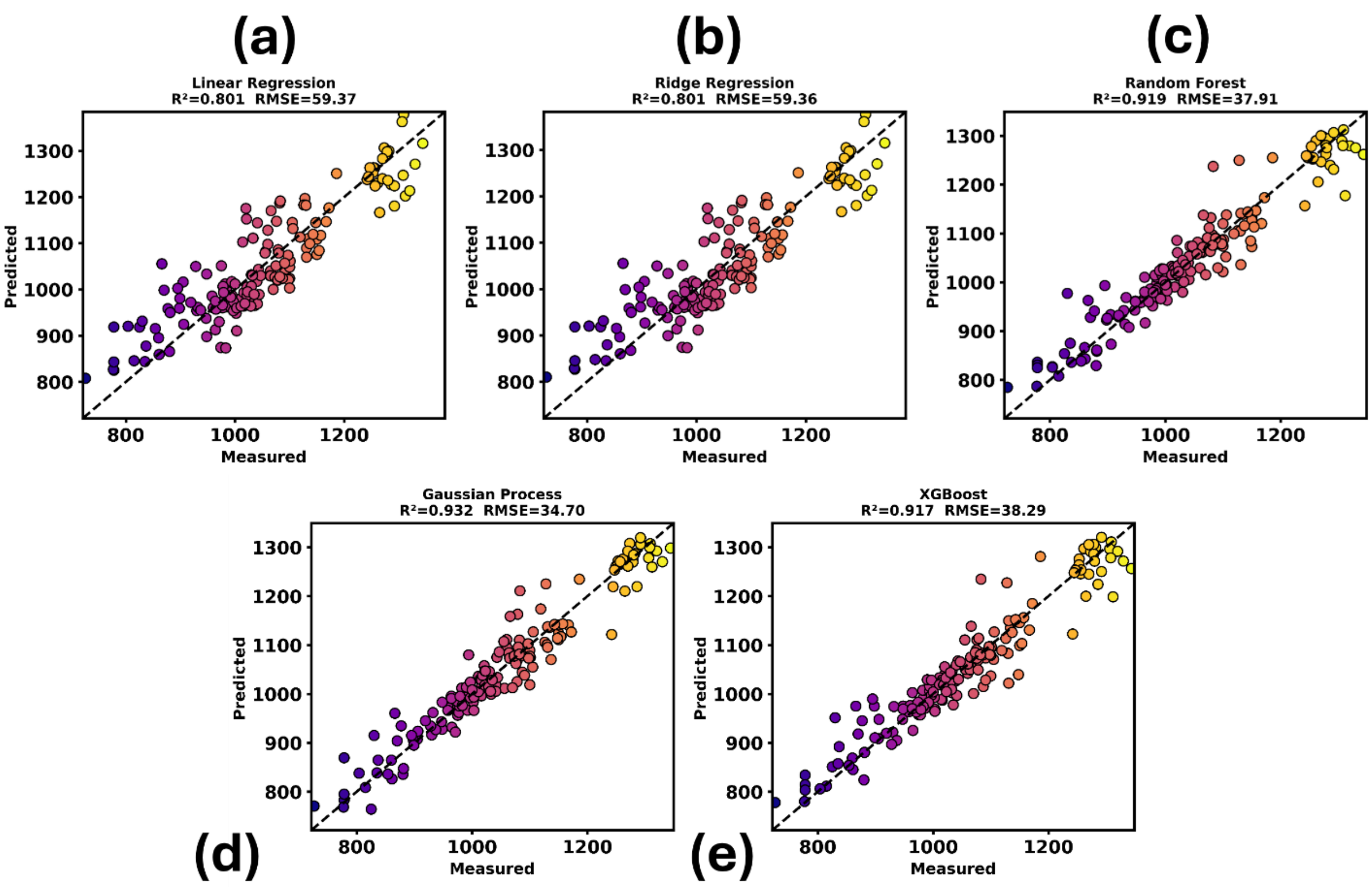

***Figure 2.*** *Leave-one-out cross-validation (LOOCV) parity plots comparing experimentally measured and predicted γ′ solvus temperatures obtained using (a) Linear Regression, (b) Ridge Regression, (c) Random Forest, (d) XGBoost, and (e) Gaussian Process Regression. The dashed line represents perfect agreement between predicted and experimental values.*

The predictive capability of different regression algorithms was evaluated using LOOCV, and the parity plots are shown in Fig. 2. Linear Regression and Ridge Regression show considerable variability around the optimal prediction line, reflecting their restricted capability to model the nonlinear associations between alloy characteristics and γ′ solvus temperature. Both linear models attained the same coefficient of determination ($R^2 \approx 0.80$), indicating that linear assumptions fall short in effectively representing the intricate descriptor-property relationship. In comparison, the nonlinear machine learning models showed significantly better alignment with experimental data. Random Forest and XGBoost produced $R^2$ values of 0.919 and 0.917, respectively, whereas the GPR reached the highest prediction accuracy with an $R^2$ of 0.932. The parity plot for the GPR model shows that most predictions closely align with the 1:1 line across the full range of $\gamma'$ solvus temperatures, suggesting very little systematic bias. The enhanced performance of the nonlinear models indicates that the relationship between alloy chemistry and the $\gamma'$ solvus temperature is fundamentally nonlinear. Intricate relationships among lattice distortion, mixing enthalpy, and electronic structure cannot be adequately captured by simple linear regression models.

### 3.3 Model benchmarking and selection

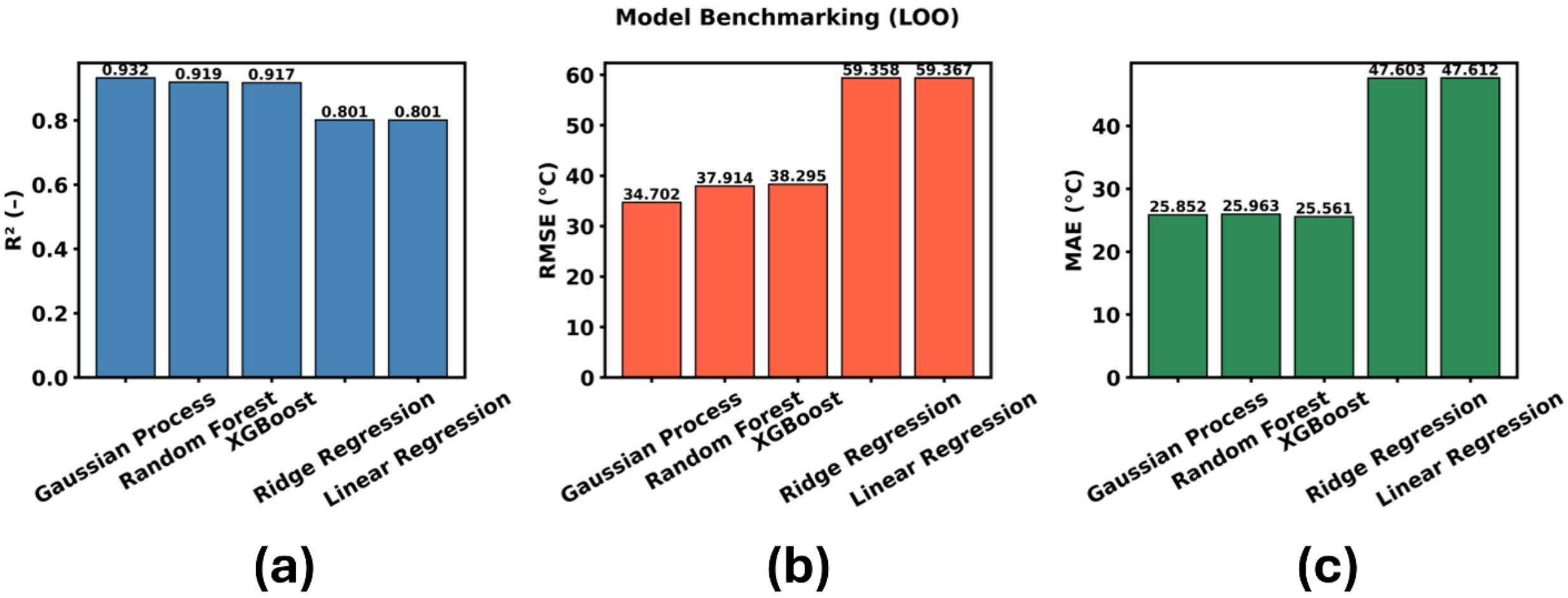

***Figure 3.*** *Performance comparison of the evaluated machine learning models. (a) Coefficient of determination ($R^2$), (b) root mean square error (RMSE), and (c) mean absolute error (MAE) obtained from leave-one-out cross-validation for Linear Regression, Ridge Regression, Random Forest, XGBoost, and Gaussian Process Regression.*

Figure 3 summarizes the quantitative comparison of model performance through the use of the coefficient of determination ($R^2$), root mean square error (RMSE) and mean absolute error (MAE). Of all the algorithms assessed, the Gaussian Process Regressor demonstrated the highest prediction accuracy, achieving an $R^2$ of 0.932, along with the lowest RMSE (34.70 °C) and MAE (25.85 °C). Random Forest and XGBoost demonstrated similar performance, achieving $R^2$ values above 0.91 and prediction errors under 40 °C, validating the efficiency of nonlinear ensemble learning techniques for forecasting alloy properties. The significantly inferior outcomes of Linear Regression and Ridge Regression (RMSE ≈ 59 °C and MAE ≈ 48 °C) further demonstrate that the $\gamma'$ solvus temperature cannot be adequately represented by linear descriptor combinations alone.

The excellent performance of the Gaussian Process model is especially beneficial because, unlike traditional regression methods, GPR simultaneously predicts both the expected $\gamma'$ solvus temperature and its associated predictive uncertainty. This probabilistic ability is crucial for Bayesian optimization, as uncertainty data directs the investigation of uncharted areas of composition space while preventing excessive extrapolation. As a result, the Gaussian Process Regressor was chosen as the surrogate model for later inverse alloy design and optimization.

**3.4 Multi-objective optimization using NSGA-II**

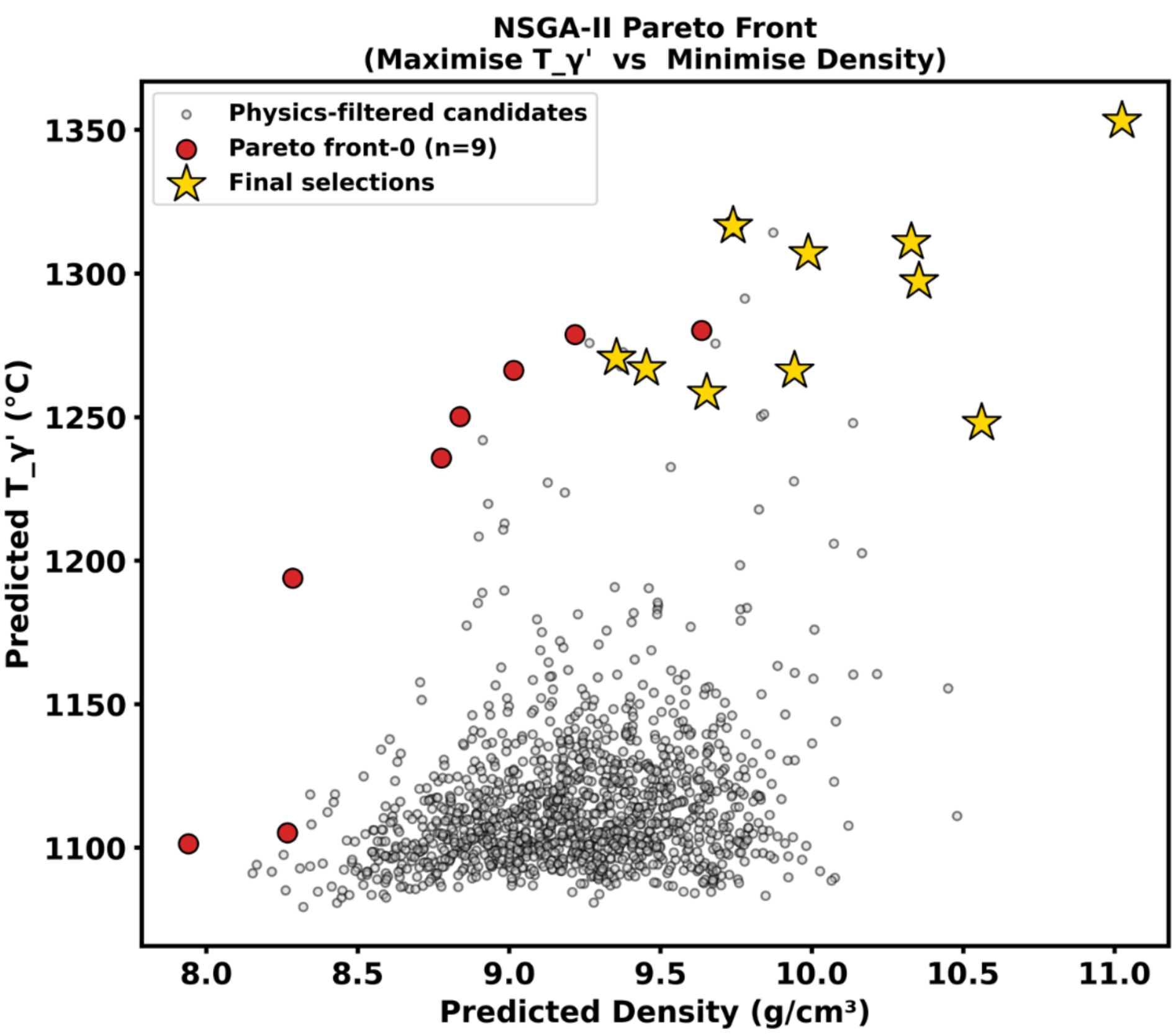

***Figure 4.*** *Multi-objective optimization using NSGA-II for simultaneous maximization of $\gamma'$ solvus temperature and minimization of alloy density. Gray markers represent physics-filtered candidate alloys, red markers denote the first non-dominated Pareto front, and gold stars indicate the selected optimized alloy compositions.*

The results of the multi-objective optimization are presented in Fig. 4, where the predicted $\gamma'$ solvus temperature is maximized simultaneously with minimizing alloy density using the NSGA-II algorithm. Before optimization, the candidate space was constrained using physics-informed descriptor limits derived from experimentally validated high-performance alloys, ensuring the search remained within chemically relevant regions of the design space. Most of the candidate alloys selected based on physics are grouped between 8.8–9.8 g $cm^{-3}$, with predicted γ′ solvus temperatures around 1150 °C, suggesting that while these compositions meet the descriptor requirements, they do not effectively optimize both goals simultaneously. Only a small selection of compositions comprises the initial non-dominated Pareto front (Front-0), as shown by the red markers in Fig. 4. These alloys achieve the ideal balance between thermal stability and density, where improving one goal cannot be achieved without sacrificing the other. A distinct pattern is seen along the Pareto front. Raising the anticipated γ′ solvus temperature typically requires a higher alloy density, indicating greater inclusion of refractory alloying elements such as W and Ta, which are recognized for improving γ′ stability while also increasing mass density. In contrast, reduced thermal stability is observed in lower-density alloys due to lower concentrations of these heavier elements. Thus, the Pareto front quantitatively demonstrates the essential design trade-off between high-temperature performance and weight minimization in Co-based superalloys. The ultimate optimized alloys, marked by the star symbols, are purposefully spread across various areas of the Pareto front instead of being clustered at one extreme. This shows that the optimization framework is designed to deliver a variety of candidate alloys appropriate for various engineering needs rather than solely pinpointing the highest predicted γ′ solvus temperature. For applications where weight is critical, alloys located toward the lower-density region of the Pareto front may be preferred, whereas stationary high-temperature components may benefit from compositions possessing the highest predicted γ′ solvus temperatures.

### 3.5 Comparison of Bayesian optimization, genetic algorithm and NSGA-II candidate selection

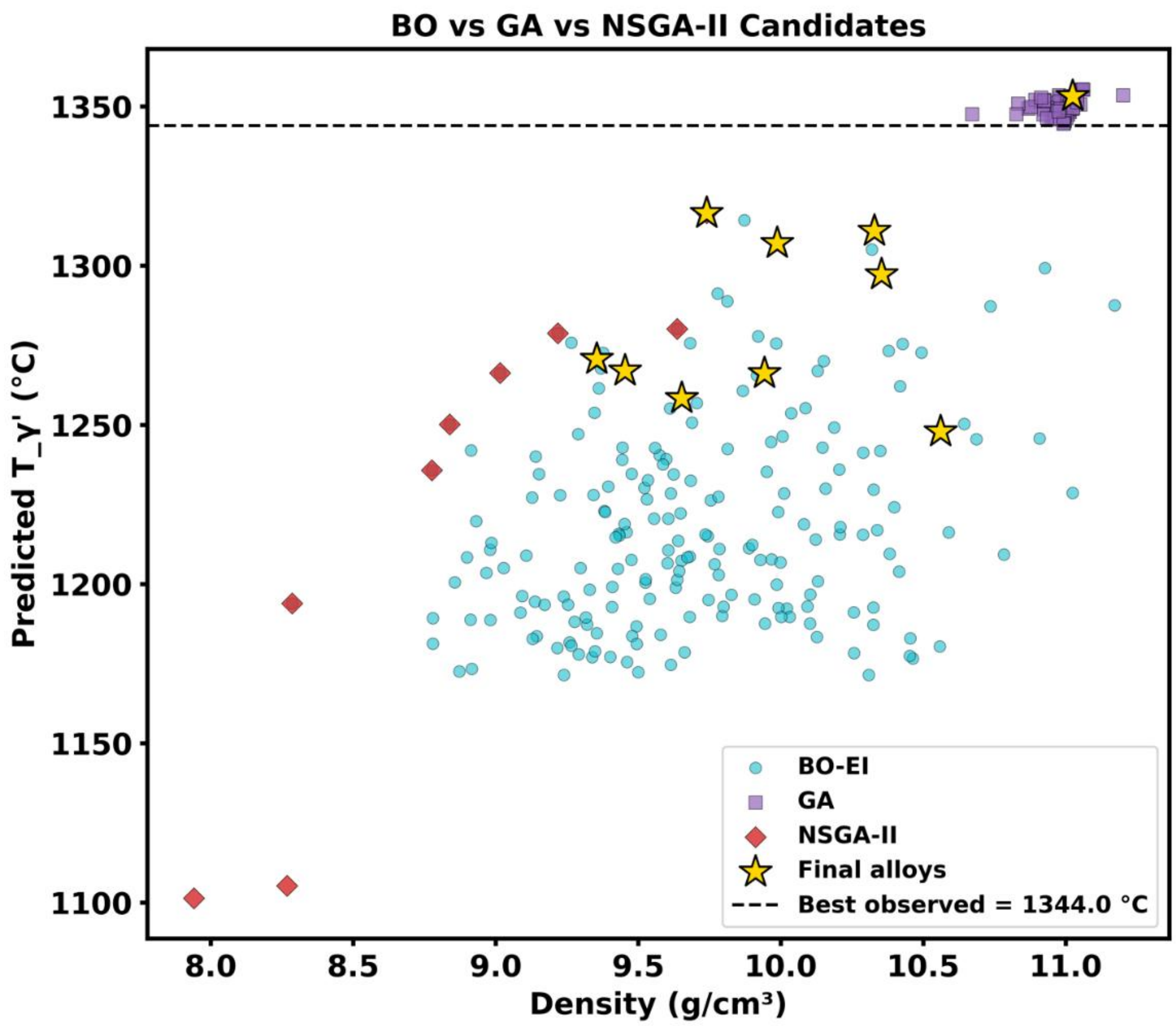

***Figure 5.*** *Comparison of candidate alloys generated using Bayesian Optimization (BO), Genetic Algorithm (GA), and NSGA-II in the density–γ′ solvus temperature design space. The dashed horizontal line indicates the highest γ′ solvus temperature reported in the experimental dataset, while the gold stars represent the final selected alloy compositions.*

The alloy candidates produced by the three optimization methods are illustrated in Fig. 5. While all methods use the same Gaussian Process surrogate model, each optimization algorithm explores the design space based on different objectives, leading to distinct candidate distributions. The Bayesian optimization candidates (cyan circles) span a wide range of densities and anticipated γ′ solvus temperatures. Bayesian optimization uses the Expected Improvement acquisition function to balance exploration and exploitation by evaluating both the expected outcome and the model's uncertainty simultaneously. As a result, the BO candidates are broadly distributed across the viable design space, enabling effective discovery of untested compositions while preventing early convergence. Conversely, the Genetic Algorithm (purple squares) quickly approaches compositions that exhibit the highest predicted γ′ solvus temperatures. The majority of GA candidates are clustered around 1345–1355 °C, indicating the effective use of the surrogate model. Nonetheless, this convergence takes place at fairly high densities (~11 g $cm^{-3}$), suggesting that optimization focused only on maximizing γ′ solvus tends to preferentially select compositions rich in refractory elements. Although these alloys demonstrate outstanding anticipated thermal stability, they could be less appealing for applications where weight is a critical factor. The NSGA-II solutions (red diamonds) are located in a clearly different area of the design space. Instead of aligning with the highest expected γ′ solvus temperatures, NSGA-II maintains a continuous array of non-dominated solutions across various density levels. This behavior indicates the goal of simultaneously increasing the γ′ solvus temperature while reducing density and illustrates the efficacy of Pareto-based optimization in producing engineering trade-offs rather than a sole optimum. The proposed ultimate alloys (gold stars) were selected by combining the benefits of the three optimization methods.

These alloys are spread throughout the upper section of the viable design space, with anticipated γ′ solvus temperatures ranging from about 1250 to 1355 °C, while maintaining densities of 9.3 to 11.0 g $cm^{-3}$. Crucially, various optimized alloys reach or slightly surpass the maximum γ′ solvus temperature found in the original experimental data (1344 °C, dashed line), suggesting that the inverse design framework can identify previously unexamined compositions with potentially enhanced high-temperature stability. Instead of relying on a single optimization method, the combined Bayesian Optimization–NSGA-II–Genetic Algorithm framework incorporates global exploration, analysis of multi-objective trade-offs, and local exploitation. Bayesian optimization adeptly explores uncertain regions of the composition space, NSGA-II determines the optimal trade-off between thermal stability and density, while the Genetic Algorithm hones high-performing solutions towards local optima. This complementary approach significantly increases the range of potential alloys while reducing the risk of settling on inferior compositions. The final candidate alloys therefore represent a balanced set of chemically feasible compositions that satisfy both thermodynamic performance requirements and practical engineering constraints, making them attractive targets for future CALPHAD calculations and experimental validation.

### 3.6 Compositional characteristics of the proposed alloys

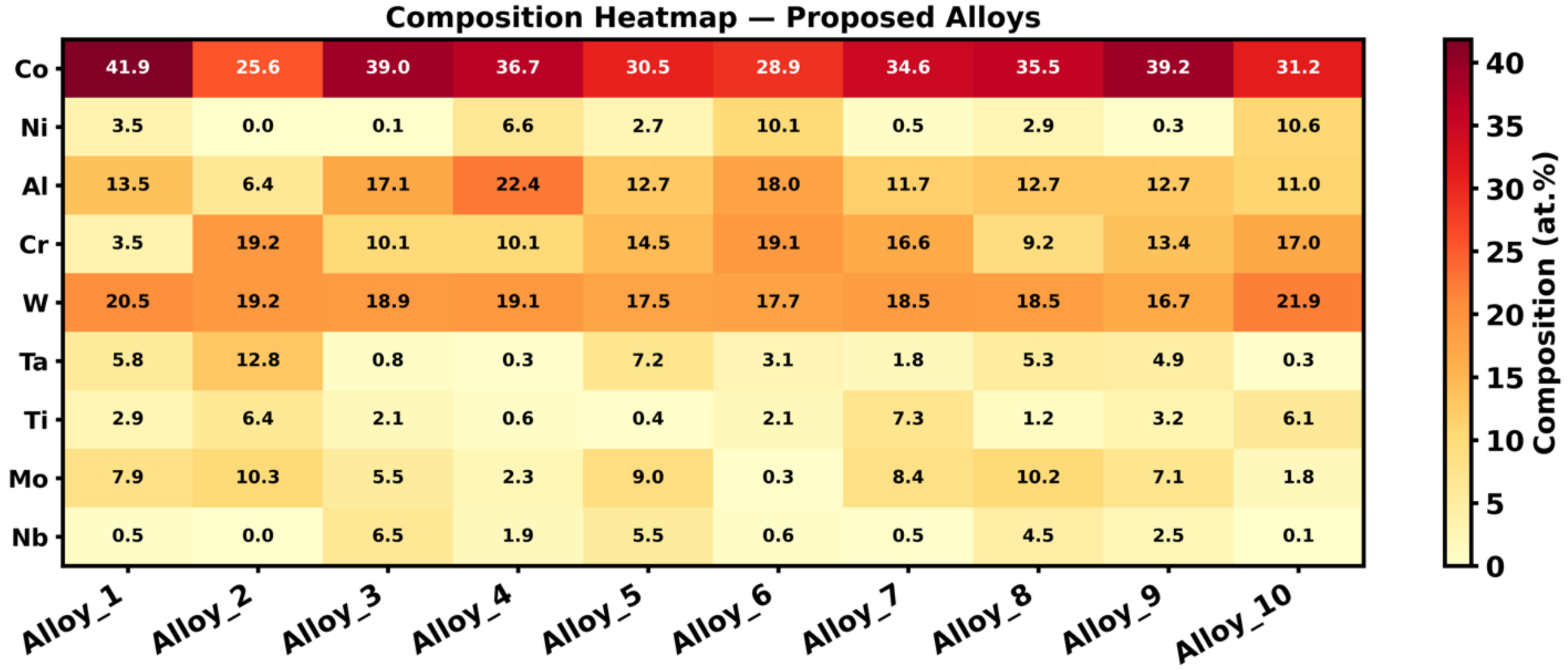

***Figure 6.*** *Heatmap showing the elemental compositions (at.%) of the ten optimized Co-based superalloys. The color scale represents the atomic percentage of each alloying element, highlighting common compositional trends and variations among the proposed alloy chemistries.*

The compositional distribution of the ten optimized Co-based superalloys is illustrated in the heatmap shown in Fig. 6. The heatmap clearly depicts the elemental design approach used by the inverse-design framework and shows multiple consistent compositional patterns among the suggested alloys. A shared feature of all optimized compositions is the preservation of Co as the primary element, with amounts varying between roughly 26–42 at.%. This illustrates the need to sustain a stable face-centered cubic $\gamma$ matrix while ensuring adequate solubility for refractory alloying elements that enhance high-temperature strength. Regardless of differences in the other alloying elements, the optimization always maintains Co as the matrix-forming element, indicating that the framework identifies compositions that align with known design principles for Co-based superalloys. Among the refractory elements, W shows notably stable concentrations (16.7–21.9 at.%) in all suggested alloys. The restricted variation indicates that tungsten is key to attaining elevated $\gamma'$ solvus temperatures. Tungsten is recognized for its ability to lower atomic diffusion, enhance lattice distortion, and boost the stability of both the $\gamma$ matrix and $\gamma'$ precipitates at high temperatures. The optimization thus gravitates towards compositions with comparatively high W additions, signifying that this element is essential for enhancing thermal stability. In comparison, Cr shows significantly greater compositional variation (3.5–19.2 at. %), indicating enhanced flexibility in meeting the optimization goals. Chromium mainly enhances resistance to oxidation and corrosion, while also affecting phase stability. The observed variation suggests that different Cr levels can achieve the desired $\gamma'$ solvus temperature through their interactions with other alloying elements. Aluminum, the main $\gamma'$-producing element, ranges from about 6.4 to 22.4 at. %. While all optimized alloys have significant Al contributions, the fairly wide composition range suggests that varying mixtures of Al and refractory elements can yield similar $\gamma'$ stability. This behavior emphasizes the collaborative effect of chemical ordering and lattice distortion on the estimated $\gamma'$ solvus temperature. The leftover refractory elements, specifically Ta, Mo, Ti, and Nb, show slight differences among the optimized alloys. These components primarily serve as secondary reinforcement additives that refine lattice mismatch, partitioning behavior, and precipitate stability. The wider compositional ranges imply that various alloy chemistries can achieve comparable expected performance by using different combinations of strengthening mechanisms rather than relying on a single composition. A significant result of the optimization is the persistently low Ni concentration, which remains below about 11 at.% across all suggested alloys and approaches zero in various compositions. This finding suggests that, within the studied descriptor space and optimization goals, significant Ni additions are not necessary for achieving the highest $\gamma'$ solvus temperature. Rather, the optimization predominantly replaces refractory alloying elements to attain improved thermal stability while preserving the Co-rich matrix. The heatmap indicates that the inverse-design framework does not settle on a single optimal chemistry but instead identifies a group of chemically diverse Co-based superalloys that exhibit similar design traits. The contents of Co and W are consistently maintained, while the levels of Cr, Al, Ta, Mo, Ti, and Nb are varied in different ratios to obtain similar $\gamma'$ solvus temperatures. This variety offers adaptability for future alloy development, enabling other factors, such as oxidation resistance, density, manufacturability, and raw material costs, to be incorporated without significantly affecting the anticipated high-temperature performance.

### 3.7 Predicted performance of the optimized alloys

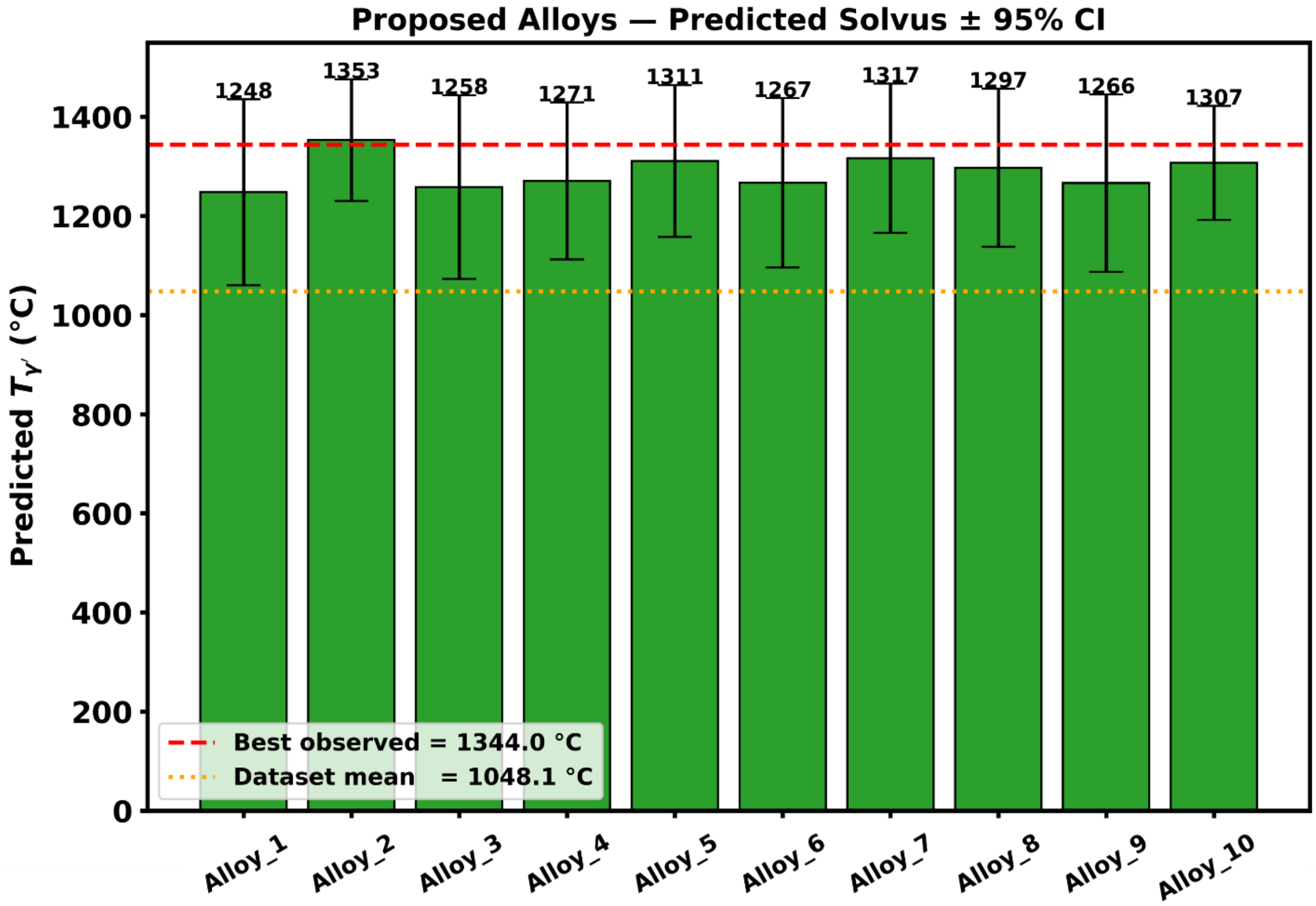

***Figure 7.*** *Predicted γ′ solvus temperatures of the optimized Co-based superalloys obtained using Gaussian Process Regression. Error bars represent the corresponding 95% confidence intervals, while the dashed horizontal line denotes the average γ′ solvus temperature of the experimental dataset.*

The predicted γ′ solvus temperatures together with the corresponding 95% confidence intervals for the ten proposed alloys are presented in Fig. 7. All proposed alloys exhibit predicted γ′ solvus temperatures substantially higher than the average value of the experimental dataset (1048 °C), indicating that the optimization framework consistently identifies compositions possessing enhanced thermal stability. The predicted solvus temperatures range from approximately 1248 °C to 1353 °C, demonstrating a significant improvement over the majority of experimentally reported alloys included in the training dataset. Among the optimized compositions, Alloy 2 exhibits the highest predicted γ′ solvus temperature of approximately 1353 °C, marginally exceeding the highest experimentally observed value (1344 °C). Several additional alloys possess predicted γ′ solvus temperatures exceeding 1300 °C, suggesting that multiple chemically distinct compositions may achieve comparable high-temperature stability rather than a single optimum alloy. The confidence intervals generated by the Gaussian Process model provide valuable information regarding prediction reliability. Although the uncertainty varies among the proposed alloys, all confidence intervals remain relatively narrow compared with the overall range of $\gamma'$ solvus temperatures within the dataset, indicating stable model predictions throughout the selected design space. The inclusion of predictive uncertainty further enables future experimental validation efforts to prioritize candidate alloys that exhibit both high predicted performance and relatively low uncertainty. Collectively, these results demonstrate that the integrated inverse-design framework successfully identifies multiple chemically feasible Co-based superalloys with predicted thermal stability comparable to or exceeding the best-performing compositions currently available in the experimental database. The existence of several high-performing candidates also provides flexibility for subsequent optimization with respect to additional engineering criteria such as density, oxidation resistance, cost, or phase stability.

### 3.8 Principal component analysis (PCA) of proposed alloys

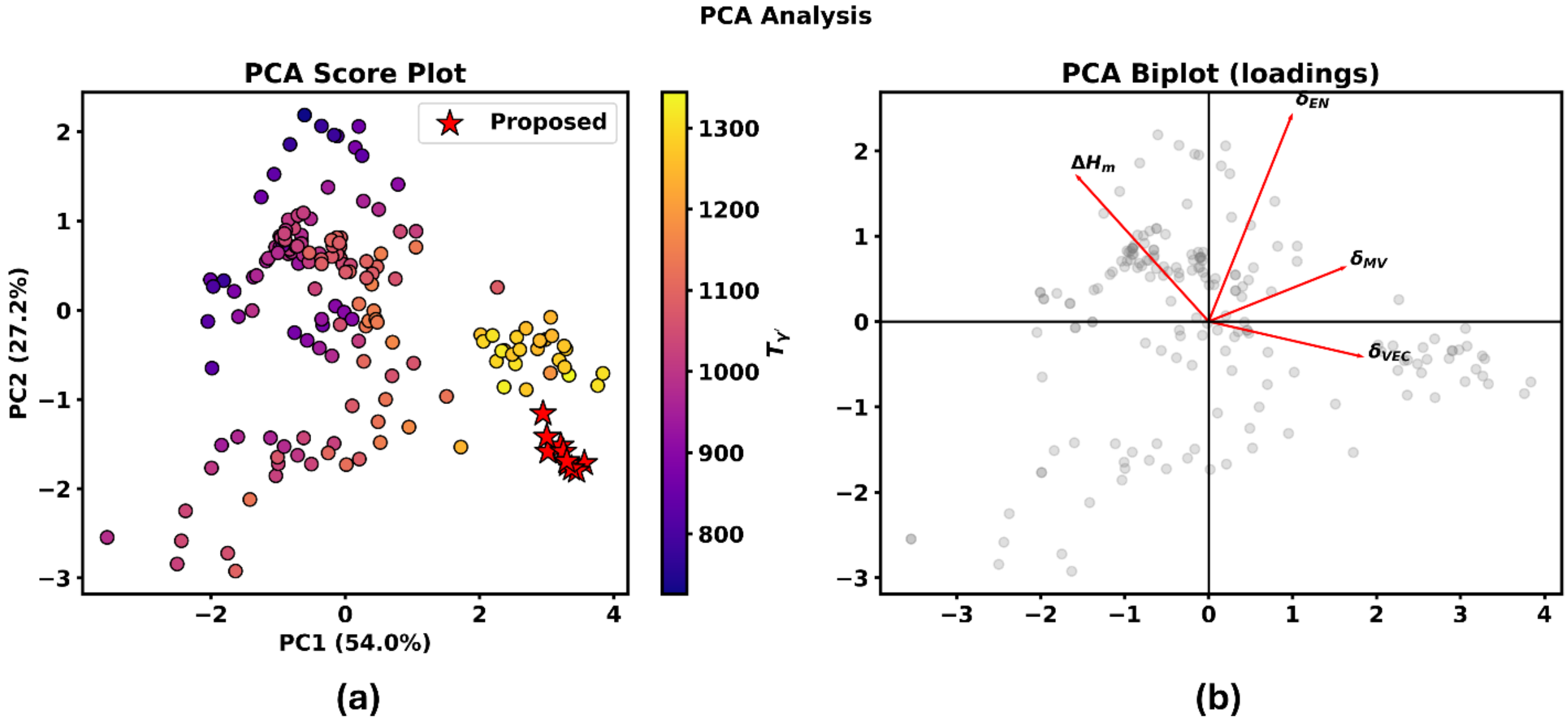

***Figure 8.*** *PCA of the experimental and optimized Co-based superalloys based on the four physics-informed descriptors. (a) PCA score plot showing the distribution of the experimental alloy compositions colored according to γ′ solvus temperature ($T_{\gamma'}$), with the optimized alloy compositions highlighted by red stars. (b) PCA loading plot illustrating the contributions of atomic size mismatch (δMV), mixing enthalpy (ΔHm), electronegativity mismatch (δEN), and valence electron concentration mismatch (δVEC) to the first two principal components, highlighting the descriptor relationships governing γ′ phase stability.*

PCA was conducted to examine the positioning of the optimized alloys relative to the experimental dataset in the reduced descriptor space. Fig. 8 displays the PCA score plot and loading plot. The first two principal components account for approximately 81% of the overall variance (PC1 = 54.0% and PC2 = 27.2%), suggesting that most of the descriptor information is preserved in the two-dimensional projection. The PCA score plot (Fig. 8a) shows that the suggested alloys group within the high-performance area of the descriptor space, staying close to the experimental data distribution. Significantly, the optimized alloys are located close to, rather than far beyond, the experimental domain, indicating that the inverse design framework performs controlled extrapolation rather than producing chemically implausible compositions. This observation enhances confidence in the dependability of the suggested alloy chemistries. The loading plot (Fig. 8b) demonstrates the influence of each descriptor on the principal components. The vectors associated with δMV and δVEC primarily align with the positive PC1 direction, suggesting that these features play a crucial role in the dataset's main variance. In contrast, ΔHm is aligned in the opposite direction, indicating an inverse relationship with the structural descriptors. The δEN vector is mainly oriented with PC2, indicating that the electronegativity difference significantly influences secondary changes in alloy chemistry. The relative positioning of the descriptor vectors additionally uncovers relationships among the physicochemical descriptors. The parallel trends of δMV and δVEC show a positive correlation between lattice distortion and electronic structure, while the contrasting direction of ΔHm implies that more negative mixing enthalpy is associated with greater lattice distortion in high-performance alloys. These relationships align with the previously discussed descriptor–property analysis and validate that the optimized alloys reside in physically significant areas of descriptor space.

### 3.9 Descriptor importance analysis

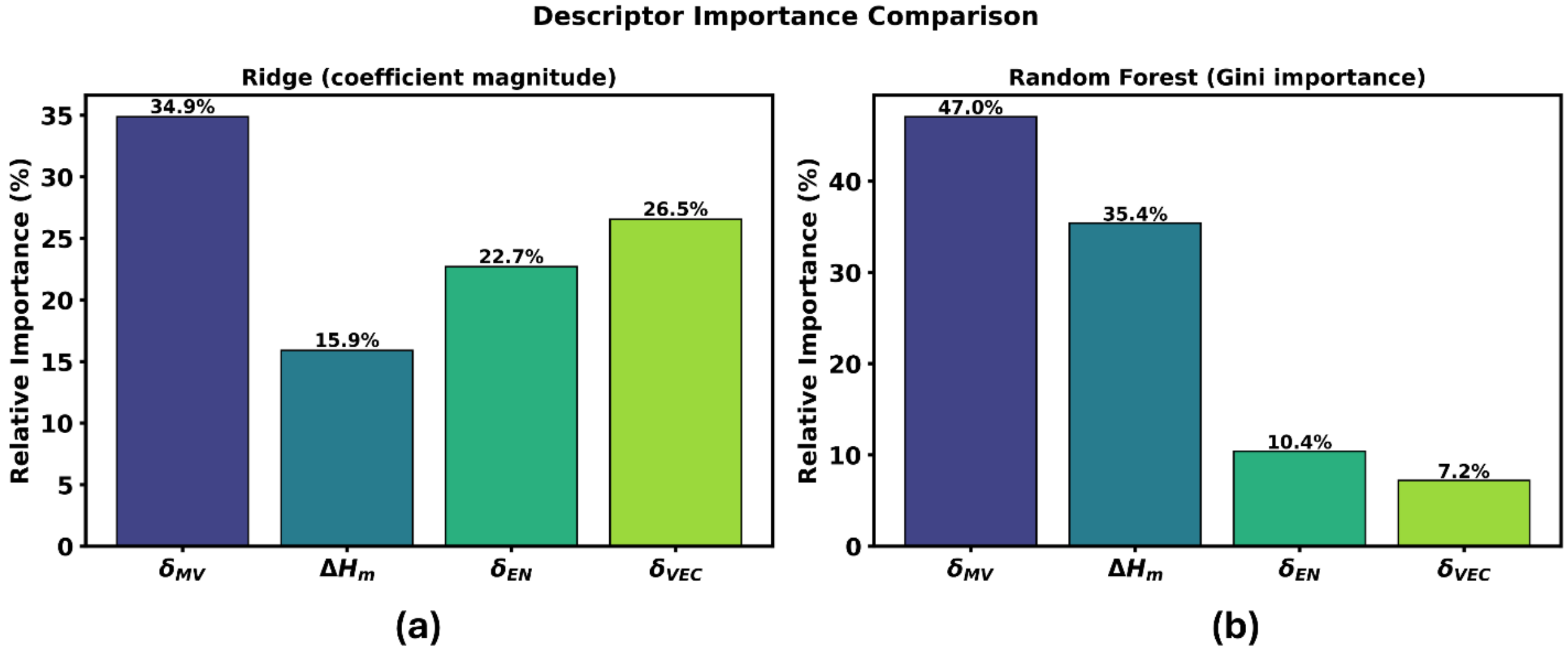

***Figure 9.*** *Relative importance of the selected physics-informed descriptors for predicting γ' solvus temperature. (a) Ridge Regression coefficient importance and (b) Random Forest feature importance illustrating the contributions of atomic size mismatch (δMV), mixing enthalpy (ΔHm), electronegativity mismatch (δEN), and valence electron concentration mismatch (δVEC).*

To assess the relative impact of the chosen physicochemical descriptors on the prediction of γ′ solvus temperature, the importance of descriptors was analyzed using Ridge regression coefficients and Random Forest Gini importance, as illustrated in Fig. 9. Utilizing two separate methods allows for evaluation of descriptor importance from both linear and nonlinear viewpoints. The Ridge regression analysis shows that the molar volume mismatch (δMV) is the primary descriptor, accounting for around 34.9% of the total model significance, succeeded by the valence electron concentration mismatch (δVEC, 26.5%), electronegativity mismatch (δEN, 22.7%), and mixing enthalpy (ΔHm, 15.9%). Given that Ridge regression assumes linear relationships, these findings primarily indicate the direct effect of each descriptor on the estimated γ′ solvus temperature. Using the Random Forest model reveals a different trend: δMV accounts for about 47% of the overall importance, with ΔHm at 35.4%, while δEN (10.4%) and δVEC (7.2%) have significantly lower contributions. The increased emphasis on ΔHm in the nonlinear model indicates that chemical interactions become more relevant when nonlinear coupling between descriptors is considered. The consistent top ranking of δMV across both methods reinforces the conclusion of ***Section 3.1*** that atomic-size mismatch is the single most influential physical driver of γ′ stability captured by this descriptor set. This finding is independently corroborated by Liao et al. [38], who identified δMV and ΔHm as the two dominant SHAP-ranked features governing $T_{\gamma'}$ across a broader Co-base/Co-Ni-base/Ni-base dataset and confirmed a near-linear combined dependence of $T_{\gamma'}$ on these two descriptors following experimental synthesis of nine extrapolated alloys. The present analysis extends this observation by showing that the same two-descriptor dominance persists even when two additional descriptors (δEN, δVEC) and both a linear (Ridge) and a nonlinear (Random Forest) importance metric are introduced, suggesting that the δMV/ΔHm-dominated behaviour is a robust feature of γ′-strengthened superalloy chemistry rather than an artifact of a particular feature set or model choice.

Increased lattice distortion promotes solid-solution strengthening and alters the elemental distribution between the γ and γ′ phases, thereby improving precipitate stability at high temperatures. Likewise, the notable role of ΔHm highlights the significance of chemical bonds in stabilizing structured intermetallic phases. The reduced individual significance of δEN and δVEC does not mean these descriptors are insignificant; instead, they mainly add value by interacting with other descriptors rather than functioning on their own. In general, the alignment between the two importance analyses suggests that the chosen descriptors effectively capture the key physical processes influencing the γ′ solvus temperature, thereby affirming their applicability for machine-learning-supported alloy design.

### 3.10 Gaussian Process surrogate model and optimization landscape

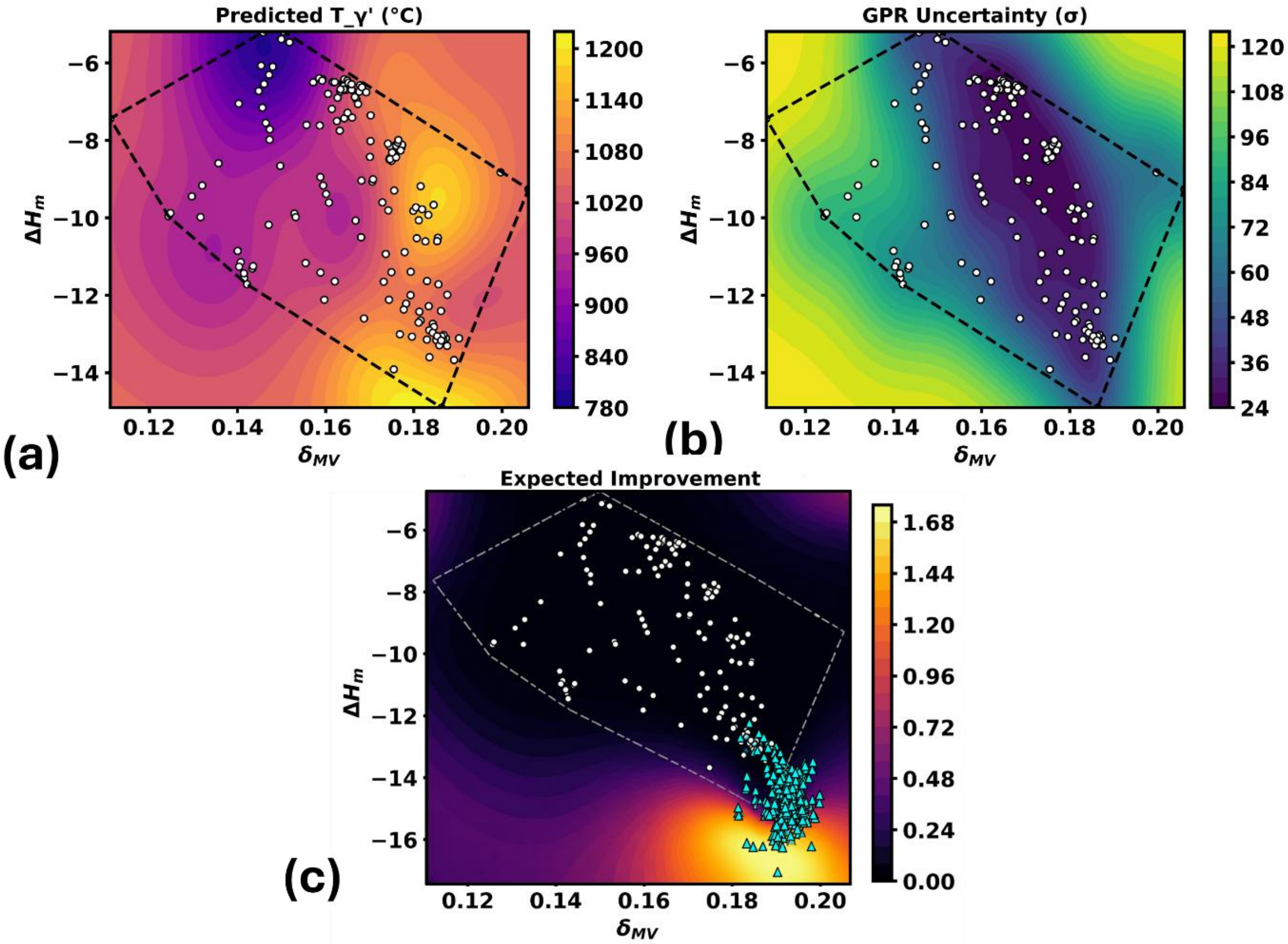

***Figure 10.*** *Gaussian Process Regression surrogate model used for Bayesian optimization. (a) Predicted γ' solvus temperature landscape, (b) predictive uncertainty (standard deviation), and (c) Expected Improvement (EI) acquisition function illustrating the regions of the descriptor space most favorable for inverse alloy design.*

The Gaussian Process surrogate model was used to predict the γ′ solvus temperature across the unexplored composition space and to quantify prediction uncertainty. Figure 10 displays the resulting prediction landscape, the uncertainty distribution, and the Expected Improvement (EI) acquisition function. The anticipated γ′ solvus surface (Fig. 10a) indicates a significantly nonlinear correlation between the descriptor space and alloy performance. Areas with significant molar volume mismatch (δMV) and markedly negative mixing enthalpy (ΔHm) show the highest anticipated γ′ solvus temperatures, suggesting that the concurrent optimization of lattice distortion and chemical bonding enhances the thermal stability of the γ′ phase. The related uncertainty map (Fig. 8b) indicates that prediction uncertainty is minimal in areas with abundant experimental data and increases gradually in less-sampled regions. This behavior is typical of Gaussian Process Regression and indicates that the surrogate model effectively estimates its confidence across the descriptor space. This type of uncertainty estimation is crucial for inverse alloy design, as it prevents overreaching into areas where the model's predictive power is limited. The Expected Improvement landscape (Fig. 8c) merges anticipated performance and predictive uncertainty to pinpoint the most promising candidate alloys. Instead of choosing compositions solely based on the maximum predicted $\gamma'$ solvus temperature, the acquisition function emphasizes areas that exhibit both high predicted performance and substantial potential for improvement over the currently best-performing alloy. As a result, the alloys selected via Bayesian optimization are clustered around the EI peak, providing an effective balance between exploring untested compositions and exploiting areas expected to host high-performance alloys. These findings show that the Gaussian Process surrogate not only delivers precise property predictions but also provides a useful structure for directing the search toward chemically relevant regions of the design space, while reducing unnecessary exploration.

### 3.11 Residual diagnostics of the Gaussian Process Regression model

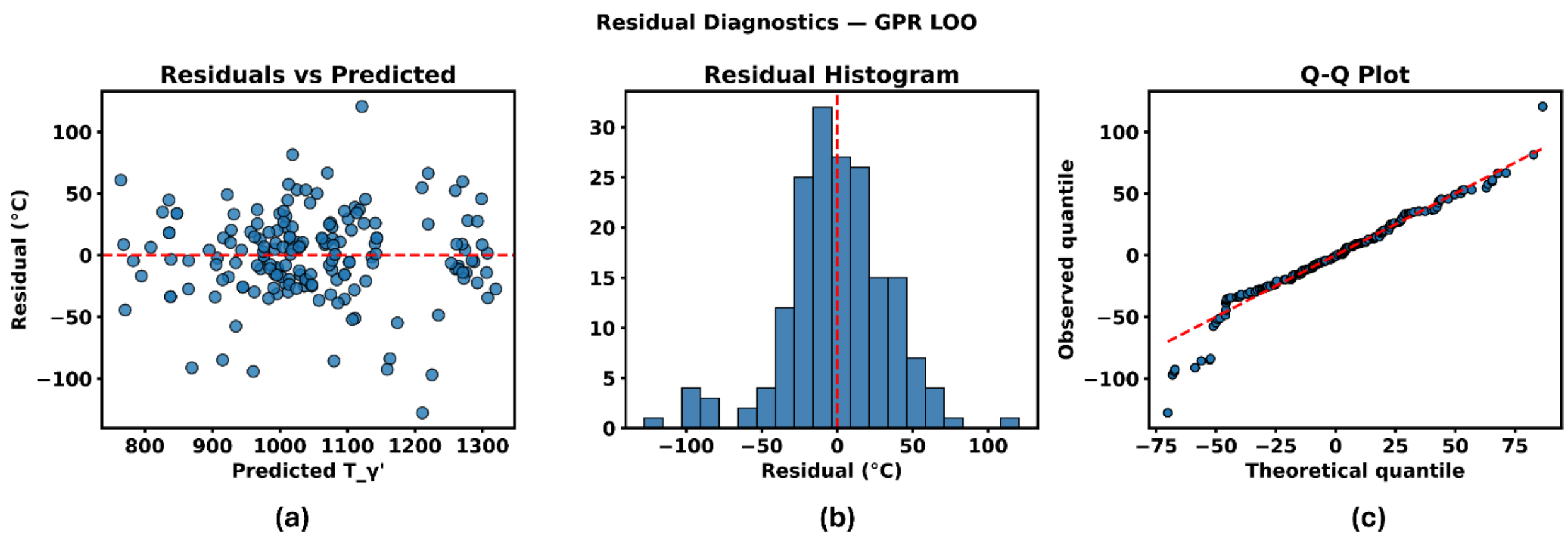

***Figure 11.*** *Residual diagnostics of the Gaussian Process Regression model. (a) Residuals versus predicted $\gamma'$ solvus temperature, (b) residual histogram, and (c) normal Q–Q plot demonstrating the statistical adequacy of the surrogate model.*

The Gaussian Process Regression model's statistical performance was additionally assessed using residual diagnostics, as shown in Fig. 11. The residual plots provide an evaluation of prediction bias, error distribution, and the model's suitability after leave-one-out cross-validation. The residuals shown against the predicted γ′ solvus temperature (Fig. 11a) are evenly spread around the zero-error line, lacking any noticeable systematic trend or curvature. The lack of organized patterns suggests that the model does not exhibit significant heteroscedasticity or consistent over- or under-prediction across the examined temperature range. The majority of prediction errors stay within roughly ±50 °C, with only a handful of larger residuals associated with compositions near the edges of the experimental design space. This type of behavior is expected for surrogate models used in intricate multicomponent alloy systems and indicates that the prediction error is largely random rather than driven by the model. The residual histogram (Fig. 11b) shows a nearly symmetric bell-shaped distribution centered on zero residual, suggesting that positive and negative prediction errors are balanced. The lack of significant skewness indicates that the model does not show a consistent tendency to either overestimate or underestimate the γ′ solvus temperature. Additionally, the fairly limited distribution indicates that most predictions are associated with small errors, consistent with the low RMSE and MAE observed during leave-one-out validation. The typical Q–Q plot (Fig. 11c) additionally validates the statistical properties of the residuals. The majority of data points closely align with the reference line, suggesting that the residuals are roughly normally distributed. Slight variations can be seen only at the extreme ends of the distribution, indicating a limited number of significant prediction errors related to compositions with either very high or very low γ′ solvus temperatures. Such variations frequently occur in materials datasets where the experimental database has a limited number of extreme compositions and do not greatly impact the model's overall predictive ability. Overall, the residual diagnostics indicate that the Gaussian Process Regression model provides unbiased predictions, with residuals roughly normally distributed and no notable systematic errors, affirming its appropriateness as a surrogate model for future Bayesian optimization and inverse alloy design.

### 3.12 Comparison with Existing Computational Alloy Design Frameworks

This study shows that integrating physics-informed descriptors with uncertainty-aware machine learning and multi-objective optimization offers an efficient approach for the inverse design of γ′-strengthened Co-based superalloys. While separate elements of the suggested framework have been investigated earlier, their integration into a cohesive workflow facilitates swift compositional evaluation, ensuring physical interpretability and explicitly addressing conflicting design goals. In contrast to traditional CALPHAD-based alloy design, which has been extensively used for Ni- and Co-based superalloys [8,39,40], the current method significantly lessens the computational workload needed to investigate high-dimensional compositional spaces. Instead of performing thermodynamic calculations for each candidate composition, the GPR surrogate quickly estimates the $\gamma'$ solvus temperature from experimental data, enabling efficient screening of 50,000 potential alloys. The suggested framework thus acts as a supplement to CALPHAD, offering an economical pre-screening phase before in-depth thermodynamic evaluation and experimental confirmation.

Earlier machine-learning research has shown the capability of surrogate models to speed up the design of Co-based superalloys [9,12,13,38,41]. Nonetheless, numerous current methods rely on empirical compositional characteristics or high-dimensional representations that offer limited physical interpretation. Conversely, the current framework consists solely of four analytically calculable descriptors—molar-volume mismatch ($\delta$MV), mixing enthalpy ($\Delta$Hm), electronegativity mismatch ($\delta$EN), and valence-electron concentration mismatch ($\delta$VEC). The correlation analysis, descriptor importance analysis, and SHAP interpretation consistently identify $\delta$MV and $\Delta$Hm as the main factors influencing the $\gamma'$ solvus temperature, supporting earlier findings [38], while indicating that $\delta$EN and $\delta$VEC provide additional predictive insights into electronic effects. Another aspect of this study is the clear validation of the predictive uncertainty linked to the GPR surrogate before Bayesian Optimization. As the Expected Improvement acquisition function is directly influenced by predictive uncertainty, uncertainty calibration provides additional assurance that the optimization process is guided by statistically reliable predictions. Additionally, in contrast to several earlier optimization studies that merge multiple objectives into a weighted objective function [42–46], this study uses NSGA-II to directly address the trade-off between the $\gamma'$ solvus temperature and alloy density via a Pareto front. This provides a set of optimal solutions that can be selected based on specific engineering needs, rather than a single composition determined by predetermined weighting factors.

Ultimately, the integration of Bayesian Optimization, NSGA-II, and a surrogate-assisted Genetic Algorithm, along with physics-informed descriptor filtering and k-means clustering, produces chemically varied candidate alloys while preventing convergence to compositionally akin solutions. The ten optimized compositions demonstrate that combining physically interpretable descriptors, uncertainty-aware machine learning, and multi-objective optimization creates an effective and reliable framework for accelerating the discovery of high-performance Co-based superalloys. These candidates represent encouraging foundations for upcoming CALPHAD computations, experimental synthesis, and assessments of high-temperature performance.

## 4. Limitations and Future Perspectives

The suggested framework shows the potential to combine physics-informed descriptors, machine learning, and multi-objective optimization to expedite alloy discovery, yet various limitations must be recognized. Initially, the optimized alloy compositions are computational predictions and have not yet been synthesized or characterized experimentally. As a result, the anticipated $\gamma'$ solvus temperatures and densities should be considered highly confident design hypotheses rather than experimentally confirmed material characteristics. Validation through alloy production, heat treatment, microstructural analysis, and differential scanning calorimetry (DSC) or high-temperature diffraction assessments will be crucial to verify the anticipated phase stability. Secondly, the predictive models were created utilizing the existing experimental database, which, while indicative of documented Co-based superalloys, does not comprehensively cover the entire multicomponent composition range. As with any data-based surrogate model, the reliability of predictions is expected to decline for compositions outside the training domain. Nonetheless, integrating physics-informed descriptors, uncertainty-aware Gaussian Process Regression, and principal component analysis helps keep the proposed alloys within chemically relevant regions of the descriptor space, thereby minimizing the risk of unrealistic extrapolation. Ultimately, the current optimization focused on $\gamma'$ solvus temperature and density as the main design goals. Other significant technological properties, such as creep resistance, oxidation behavior, hot corrosion resistance, lattice misfit, $\gamma/\gamma'$ volume fraction, phase equilibria, processing attributes, and alloy cost, were not directly included in the optimization. Expanding the framework into a genuine multi-property design approach by incorporating CALPHAD calculations, first-principles simulations, and experimental insights via active learning signifies a promising avenue for future efforts. Despite these constraints, the current research develops a robust, interpretable inverse-design approach that significantly narrows the compositional search space and provides experimentally viable Co-based superalloy candidates for future validation and advancement.

## 5. Conclusion

A physics-informed machine learning framework was developed for the inverse design of Co-based superalloys with an improved γ′ solvus temperature, combining descriptor engineering, Gaussian Process Regression, Bayesian Optimization, NSGA-II, and Genetic Algorithms. The key findings are outlined below:

- Four significant descriptors, specifically atomic size mismatch (δMV), mixing enthalpy (ΔHm), electronegativity mismatch (δEN), and valence electron concentration mismatch (δVEC), effectively represented the key physicochemical factors influencing γ′ solvus temperature. Descriptor correlation and feature importance analyses revealed δMV and ΔHm as the primary variables influencing alloy performance.
- Among the assessed regression algorithms, Gaussian Process Regression achieved the highest predictive performance, attaining an $R^2$ of 0.932, an RMSE of 34.7 °C, and an MAE of 25.9 °C using leave-one-out cross-validation. The probabilistic forecasts provided reliable uncertainty assessments that facilitated effective investigation of uncharted compositional space.
- The combination of Bayesian Optimization, NSGA-II, and Genetic Algorithms facilitated a synergistic exploration and exploitation of the design space. Bayesian Optimization effectively sampled regions with potential; NSGA-II produced optimal compromises between γ′ solvus temperature and density; and the Genetic Algorithm improved high-performing candidate compositions.
- The inverse design framework identified 10 chemically viable Co-based superalloys with predicted γ′ solvus temperatures ranging from 1248 to 1353 °C, with multiple candidates approaching or exceeding the highest experimentally recorded value in the training dataset. Principal component analysis confirmed that these optimized alloys remain close to the experimental descriptor space, suggesting physically meaningful extrapolation.
- The optimized alloys consistently maintained elevated levels of Co and W while allowing flexibility in the concentrations of Cr, Al, Ta, Mo, Ti, and Nb, indicating that various alloy compositions can attain similar high-temperature stability through different arrangements of lattice distortions, chemical interactions, and electronic structures.

The current study shows that integrating physics-driven descriptors with machine learning and multi-objective optimization yields an effective, understandable framework for accelerating the development of Co-based superalloys. The suggested approach significantly decreases the compositional search space and provides experimentally viable candidate alloys for subsequent thermodynamic evaluation and experimental verification. This framework can be easily adapted to design other high-temperature structural materials that require the concurrent optimization of competing characteristics.

**Conflict of Interest Statement**

The authors declare that they have no known competing financial interests or personal relationships that could have influenced the work reported in this paper.

**Credit Authorship**

Prashil S. Joshi: Conceptualization, Methodology, Visualization, Investigation, Writing – original draft, reviewing, Resources.

**Data Availability Statement**

The primary experimental dataset used to train and validate the surrogate models in this study was compiled from the literature, including the dataset of [38], publicly available at https://github.com/materials-informatics-nwpu/unsupervised_learning_based_sampling. Additional data supporting the findings of this study can be made available upon reasonable request.